\documentclass[reprint,aps,prl,superscriptaddress]{revtex4-2}
\usepackage{bm,physics,amsmath,amssymb,
	graphicx,xcolor,embedfile,comment}
\usepackage[utf8]{inputenc}
\usepackage[normalem]{ulem}

\usepackage[T1]{fontenc}
\usepackage{tikz}

\graphicspath{{figures/}}

\makeindex

\begin{document}


\title{Spin-Electric Coupling in Lead Halide Perovskites}

\author{Artem G. Volosniev}
\affiliation{Institute of Science and Technology Austria,
    Am Campus 1, 3400 Klosterneuburg, Austria}
    
\author{Abhishek Shiva Kumar}
\affiliation{Institute of Science and Technology Austria,
    Am Campus 1, 3400 Klosterneuburg, Austria}

\author{Dusan Lorenc}
\affiliation{Institute of Science and Technology Austria,
    Am Campus 1, 3400 Klosterneuburg, Austria}
    
\author{Younes Ashourishokri}
\affiliation{Institute of Science and Technology Austria,
    Am Campus 1, 3400 Klosterneuburg, Austria}
    
\author{Ayan A. Zhumekenov}
\affiliation{KAUST Catalysis Center (KCC), Division of Physical Sciences and Engineering, King Abdullah University of Science and Technology (KAUST), Thuwal 23955-6900, Kingdom of Saudi Arabia} 
    
 \author{Osman M. Bakr}
\affiliation{KAUST Catalysis Center (KCC), Division of Physical Sciences and Engineering, King Abdullah University of Science and Technology (KAUST), Thuwal 23955-6900, Kingdom of Saudi Arabia}    
    
\author{Mikhail Lemeshko}
\affiliation{Institute of Science and Technology Austria,
    Am Campus 1, 3400 Klosterneuburg, Austria} 
    
\author{Zhanybek Alpichshev}
\email{alpishev@ist.ac.at}
\affiliation{Institute of Science and Technology Austria,
    Am Campus 1, 3400 Klosterneuburg, Austria}

\begin{abstract}
Lead-halide perovskites enjoy a number of remarkable optoelectronic properties. To explain their origin, it is necessary to study how electromagnetic fields interact with these systems. We address this problem here by studying two classical quantities: Faraday rotation and the complex refractive index in a paradigmatic perovskite CH$_3$NH$_3$PbBr$_3$ in a broad wavelength range. We find that the minimal coupling of electromagnetic fields to the {\bf k}$\cdot${\bf p} Hamiltonian is insufficient to describe the observed data even on the qualitative level. To amend this, we demonstrate that there exists a relevant atomic-level coupling between electromagnetic fields and the spin degree of freedom. This spin-electric coupling allows for quantitative description of a number of previous as well as present experimental data. In particular, we use it here to show that the Faraday effect in lead-halide perovskites is dominated by the Zeeman splitting of the energy levels, and has a substantial beyond-Becquerel contribution. Finally, we present general symmetry-based phenomenological arguments that in the low-energy limit our effective model includes all basis coupling terms to the electromagnetic field in the linear order.
\end{abstract}

   \maketitle 

Originally, lead-halide perovskites (LHP) attracted attention as promising candidates for photovoltaic applications~\cite{Gratzel2014,Saliba2016,Shi2015}. However, it was soon realized that LHP also feature a host of other exceptional but seemingly unrelated optical properties such as efficient THz generation, high harmonic generation and even high-temperature Dicke superluminescence to name a few (see e.g.~\cite{Ferrando2018, Zhou2020, Findik2021} and references therein). This richness of physics arguably makes lead-halide perovskites unique and calls for a better understanding of what makes them so special at the microscopic level. In particular, in view of the growing consensus on the importance of Rashba-type physics in the context of these materials \cite{Zheng2015, Mosconi2017, Becker2018, Lafalce2022, Niesner2016}, there is an urge to have a unified framework for describing the interaction between electromagnetic fields and electronic degrees of freedom in lead halide perovskites.

The band structure of conduction- and valence-band electrons near the chemical potential in LHP is well understood by now. Their quantitative description can be achieved within the ${\bf k} \cdot {\bf p}$ approach based on spin-orbit-split Pb-based s- and p-orbitals hybridized with neighbouring halide s- and p-orbitals \cite{Becker2018}. To include the electromagnetic field in this description, one would naively perform the standard substitution (${\bf k} \rightarrow {\bf k} -e{\bf A}$). However, as we show below, this leads to a qualitative disagreement with experimental measurements of some classic quantities such as the linear refractive index and Faraday rotation. In this Letter, we show that, in order to achieve a quantitative agreement with experiment, it is necessary to introduce additional coupling terms into the effective model of LHP. Physically these terms are related to the atomic polarization of Pb atoms, and to the details of the electronic structure of LHP, which leads to a very specific Zeeman-type interaction. By using very general arguments, we construct a phenomenological Hamiltonian that includes both electric and magnetic fields, and fix the magnitude of each term by comparing the predictions of the theory with our measurements.

We focus on basic optical properties such as frequency-dependent absorption and the dielectric constant $\epsilon(\omega)$, which can provide important insight into the microscopic structure of a material~\cite{Green2015,Leguy2016}. We also measure the Faraday effect -- a rotation of polarization of light that propagates inside a sample in the presence of a magnetic field~\cite{Zommerfeld1954}. One important aspect of this classic phenomenon is that in order to account for it, one needs to consider the effect of both electric and magnetic fields on the material. This makes it particularly useful for understanding details of the electronic coupling to electromagnetic fields. 

{\it Faraday rotation.---} The observable that quantifies strength of the Faraday effect in a given medium is the so-called Verdet constant, $V$, which is defined via $\Theta_{F} = V\, B \, L$, where $\Theta_{F}$ is the angle between the initial and final polarizations; $B$ is the strength of the magnetic field; $L$ is the thickness of the sample. In general, $V$ is a function of frequency of the probing light. As is well known \cite{Zommerfeld1954}, $V(\omega)$ tends to diverge near absorption regions. In many cases this behaviour can be captured by the classical Becquerel formula: $V(\omega)=\gamma \frac{e}{2mc}\omega \frac{\partial n}{\partial \omega}$, with $n(\omega)$ being the (phase) refractive index and $\gamma\!\approx\!1$ is a numeric fitting parameter.

Large Verdet constants have recently been demonstrated in the vicinity of the band gap transition in methylammonium lead bromide perovskites CH$_3$NH$_3$PbBr$_3$~\cite{Sabatini2020}, suggesting industrial applications of LHP as Faraday rotators. However, the behavior of $V(\omega)$ near the energy gap (arguably) cannot provide any deep understanding of microscopic physics.  One reason is the extreme sensitivity of $V(\omega)$ to the value of the gap energy $\Delta$, as the Becquerel formula suggests; another reason is the presence of a strong exciton near the absorption edge (see Fig.~\ref{fig:fit_mu}B) which dominates $n(\omega)$ in this frequency range, obscuring the details of the underlying basic single-particle physics. 

\begin{figure}
\includegraphics[scale=0.25]{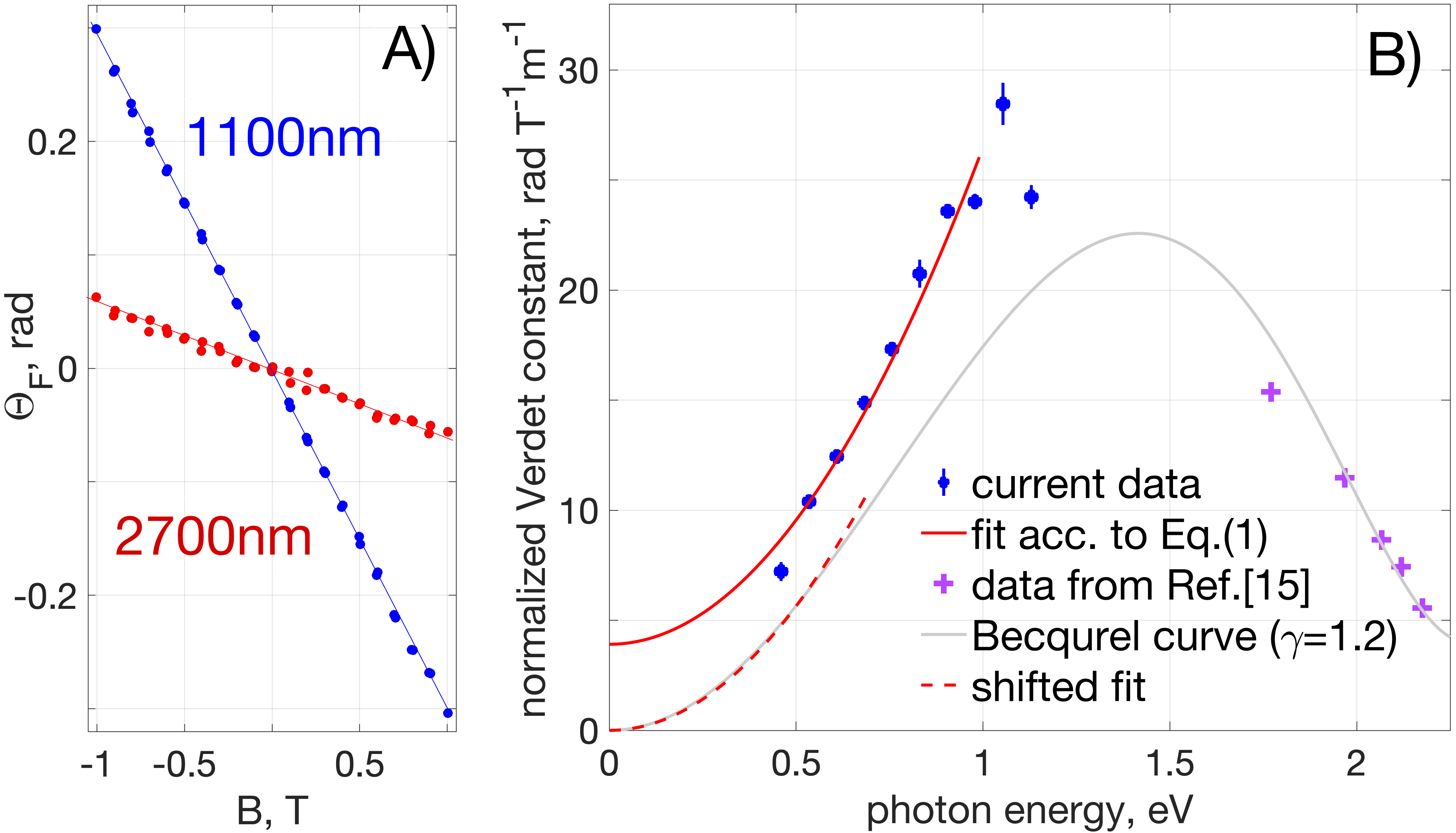}
\caption{A) Faraday rotation $\Theta_F$ as a function of magnetic field $B$ for two wavelengths; B) Normalized Verdet constant $\tilde{V}(\omega)$ extracted from the data in A) as a function of the photon energy of the incident beam (blue circles); red curve: fit according to Eq.~(\ref{eq:fit_Verdet}) (see text); purple crosses: high frequency data of Ref.~\cite{Sabatini2020} shown here for comparison; gray curve: normalized Verdet constant $\tilde{V}(\omega)$ computed from the Becquerel formula $\gamma (e/2mc)\,\omega\, {\partial n}/{\partial \omega}$ with $n(\omega)$ extracted from the data in Fig.~\ref{fig:fit_mu}B. Note that the fit according to Eq.~(\ref{eq:fit_Verdet}) differs from the Becquerel curve only by a constant shift (red dashed curve) at low frequencies.}
\label{fig:faraday}
\end{figure}

In this Letter, we study the Faraday effect in a bulk single-crystal CH$_3$NH$_3$PbBr$_3$ sample in a broad infrared range $\lambda\!=\!1100-2700$nm. The tunable-wavelength probe beam is generated by an optical parametric amplifier. Throughout the experiment the sample is kept in vacuum at $T\!=\!260$K nominal temperature (cubic phase) in an optical magnet cryostat; the field is varied between $B\!=\!\pm 1$T; the polarization rotation is detected with a pair of balanced pyroelectric IR detectors (see Fig.~\ref{fig:faraday}A). High quality bulk single crystal samples of CH$_3$NH$_3$PbBr$_3$ were grown by inverse temperature crystallization method as described elsewhere \cite{Saidaminov2015,Saidaminov2015_2}. More experimental details can be found in Supplementary Material~\cite{SM}.  

The measured Faraday rotations as a function of $B$ is presented in Fig.~\ref{fig:faraday}A. The slope of each line is proportional to the Verdet coefficient at this wavelength. It is known from general considerations that $V(\omega)$ must be an even function of frequency ~\cite{Bennett1965}, therefore suggesting the following functional form to analyze the data at low frequencies:
\begin{equation}
V(\omega)=\frac{a_0+a_2\omega^2}{(\Delta^2-\omega^2)^2},
\label{eq:fit_Verdet}
\end{equation}
where $a_0$ and $a_2$ are fitting parameters, and $\Delta$ is the band gap. The form of the denominator is motivated by the Becquerel formula \cite{Sabatini2020}. In order to focus on the non-trivial numerator of Eq.~(\ref{eq:fit_Verdet}), we plot in Fig.~\ref{fig:faraday}B the ``normalized'' Verdet constant $\tilde{V}(\omega) = V\left(\omega)\cdot(1-(\omega/\Delta)^2\right)^2$. As can be seen in Fig.~\ref{fig:faraday}B the simple curve $a_0+a_2\omega^2$ is indeed providing a reasonable fit for photon energies less than 1eV~\footnote{As will be shown below, the fact that the value of $a_0$ is finite indicates the importance of electron hopping at low frequencies}. At high frequencies, our data as well as the data of Ref.~\cite{Sabatini2020} suggest that higher-order terms should be added to Eq.~(\ref{eq:fit_Verdet}). 

\begin{figure}[t]
\includegraphics[scale=0.28]{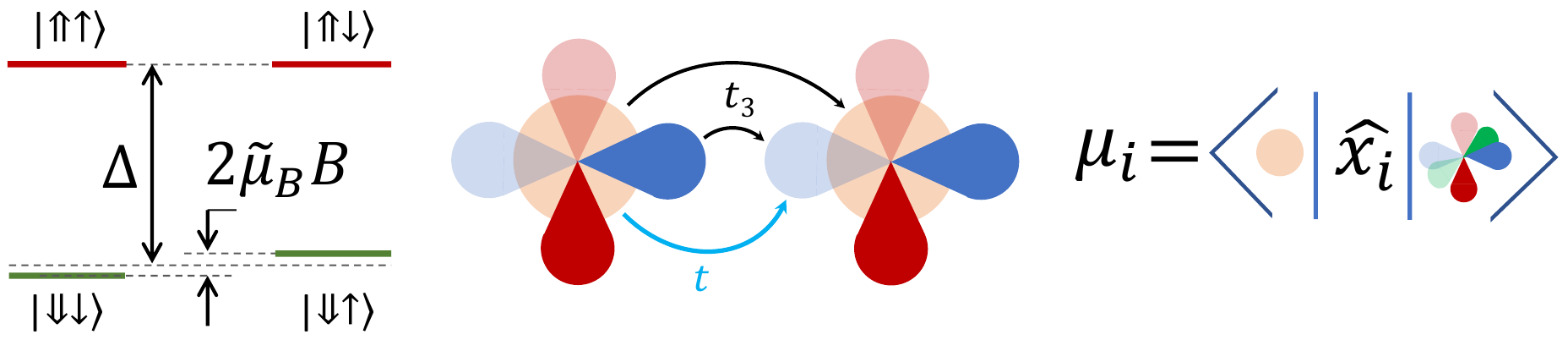}
\caption{Basic elements of the effective Hamiltonian. Left: Conduction band made of spin-orbit coupled p-type orbitals ($J\!=\!1/2$); and valence band made of s-type orbitals ($J\!=\!1/2$). Zeeman interaction projected on this basis acts only on the valence band. Middle: inter-site hopping between s- and p-orbitals; $t$ and $t_3$ correspond to the inter- and intra-orbital hoppings respectively. Right: On-site electric field-induced hybridization between s- and p-orbitals. See Eq.~(\ref{eq:states}) for the definition of the basis states.}
\label{fig:figure_states}
\end{figure}

{\it Theoretical considerations.---} Our starting point for describing the observed frequency dependence of $V(\omega)$ is the effective low-energy Hamiltonian introduced phenomenologically in~\cite{Jin2012} and in~\cite{Becker2018} via the {\bf k}$\cdot${\bf p} method~\cite{Kane1966}. This approach is based upon the fact that the low-energy optoelectronic properties of APbX$_3$ (A=Cs, CH$_3$NH$_3$; X=Cl, Br, I) can be qualitatively understood from transitions between four basis states originating from $J\!\!=\!\!1/2$  p- and s-like states mostly associated with Pb and X atoms~\cite{Umebayashi2003}. We introduce the following notation to refer to these states:
\begin{equation}
\begin{split}
 |\!\!\Uparrow\uparrow\rangle = -\left(|p_z\rangle |\!\!\uparrow\rangle +\left( |p_x\rangle +i |p_y\rangle \right) |\!\!\downarrow\rangle\right)/\sqrt{3},\\
 |\!\!\Uparrow\downarrow\rangle = \left(|p_z\rangle |\!\!\downarrow\rangle -\left( |p_x\rangle -i |p_y\rangle \right) |\!\!\downarrow\rangle\right)/\sqrt{3},\\
 |\!\!\Downarrow\uparrow\rangle =|s\rangle|\!\!\uparrow\rangle, \quad \quad \quad |\!\!\Downarrow\downarrow\rangle =|s\rangle|\!\!\downarrow\rangle.
\end{split}
\label{eq:states}
\end{equation}
Here, on the left-hand side, the thick and thin arrows correspond to the orbital (valence/conduction) and projection of the total angular momentum (quasi-spin) $J\!\!=\!\!1/2$ states, respectively. In the vicinity of the high-symmetry $R$-point (for convenience, ${\bf k}\! \equiv \! 0$), the effective Hamiltonian can be written as ($\hbar=1$):
\begin{equation}
H_k=\frac{1}{2}\left(\Delta+ t_3 ({\bf k}a)^2 \right) \tau_3 \! \otimes \!  \sigma_0 + 2 a t  \, \tau_2 \! \otimes \! \sigma_{\alpha} \, k^{\alpha},
\label{eq:naive}
\end{equation} 
\noindent where $t_3$ and $t$ are the intra- and inter-orbital hopping integrals respectively (Fig.~\ref{fig:figure_states}~middle); $a$ is the lattice constant; $\tau_{\alpha}$ and $\sigma_{\alpha}$ are the Pauli matrices acting on the orbital and quasi-spin respectively (compare with \cite{Jin2012}). In order to account for the Faraday effect, one needs to couple-in electromagnetic field into this Hamiltonian. The most obvious way to do so is to employ minimal coupling and add a Zeeman term to Eq.~(\ref{eq:naive}). However, as it turns out, in this case the normalized Verdet constant $\tilde{V}(\omega)$ is (almost) independent of the frequency contrary to our experiment data (see below).

In view of this qualitative disagreement, it is necessary to look more carefully at the interaction of electrons with electromagnetic fields. One factor naturally missing in minimal coupling to the {\bf k}$\cdot${\bf p} Hamiltonian is on-site ``atomic'' polarization due to the applied electric field. Direct calculation of the matrix elements of  $q{\vec E}\!\cdot\!{\vec r}$ over the basis Eq.~(\ref{eq:states}) produces an extra  term in the Hamiltonian (see Fig.~\ref{fig:figure_states}):
\begin{equation}H_E=\mu^{\alpha \beta} \tau_1 \! \otimes \! \sigma_{\alpha} E_{\beta} = \mu \tau_1\!\otimes\!\vec{\sigma}\!\cdot\!\vec{E},
\end{equation}
 \noindent here $q\!=\!\!-|e|$ is the charge of electron; $\mu^{\alpha \beta}= q\langle s|r^{\alpha}| p_{\beta}\rangle=\mu \delta^{\alpha \beta}$ is the (effective) atomic dipole moment; $\mu \! \sim \!q a_{\mathrm{Pb}}$ ($a_{\mathrm{Pb}}$ being the radius of Pb atom). The significance of the term $H_E$, which we dub {\it ``spin-electric''}, is that it describes the local atomic coupling of the highly polarizable Pb$^{2+}$ ions~\cite{Schwerdtfeger2019} to the external electric field $\vec{E}$.

Next, we consider magnetic coupling. Note that \!$J\!=\!1/2$ quasi-spin is not identical to the actual spin, and the Zeeman term projected on the basis in Eq.~(\ref{eq:states}) will be:
\begin{equation}
H_B=\tilde \mu_B \left( \frac{\tau_3-\tau_0}{2}\right) \!\otimes\!\vec{\sigma}\!\cdot\!\vec{B},
\end{equation}
 where $\tilde \mu_B$ is a parameter that determines the strength of the magnetic coupling.
\noindent As this form suggests, only the valence band experiences Zeeman splitting (see Fig.~\ref{fig:figure_states}).
With these considerations, our effective Hamiltonian $H=H_k+H_E+H_B$ reads as  
\begin{equation}
\begin{split}
H=&\frac{1}{2}\left(\Delta+\epsilon({{\bf k}})\right)\tau_3\!\otimes\! \sigma_0 + 2 t \, \tau_2\! \otimes \!\sigma_{\alpha} Q^{\alpha}({\bf k}) + \\
+&\mu \tau_1\!\otimes\!\vec{\sigma}\!\cdot\!\vec{E}+(\tilde \mu_B/2)(\tau_3-\tau_0)\!\otimes\!\vec{\sigma}\!\cdot\!\vec{B},
\end{split}
\label{eq:Hamiltonian}
\end{equation}
\noindent where $\epsilon({{\bf k}})$ and $Q^{\alpha}({\bf k})$ are the extensions of the corresponding functions in Eq.~(\ref{eq:naive}) across the Brillouin zone. These functions reflect the band structure; apart from general symmetry properties (see the discussion below) their exact form is not known. It can be used as a fitting factor in the present low-energy effective theory.  For the functions $\epsilon$ and $Q_\alpha$ used in Eq.~(\ref{eq:naive}), Eq.~(\ref{eq:Hamiltonian}) is rotationally symmetric, and therefore $H$ transforms according to $\Gamma_1^+$ representation of $O_h$~\cite{grouptheory}. A convenient form used in Ref.~\cite{Volosniev2022}, which also leads to the highest symmetry ($\Gamma_1^+$), is $\epsilon ({\bf k})=\sum_{\alpha=1}^3 2t_3(1-\cos(k_{\alpha}a)); \, Q_{\alpha}({\bf k})=\sin(k_{\alpha} a)$~\footnote{In this case $\epsilon$ and $Q_{\alpha}$ terms transform according to the $\Gamma_1^+$ and $\Gamma_{15}^-$ irreducible representations of the group $O_h$, correspondingly. This means that $H$ transforms according to the $\Gamma_1^+$ representation.}.

The Hamiltonian in Eq.~(\ref{eq:Hamiltonian}) allows us to calculate the low-frequency Verdet coefficient. Assuming that $t_3\! \ll\! \Delta$, we derive an expression that agrees with Eq.~(\ref{eq:fit_Verdet}):
\begin{equation} 
V=-\frac{\tilde \mu_B}{2c\epsilon_0 n a^3}\frac{4\Delta\mu^2\omega^2+C_0}{(\Delta^2-\omega^2)^2}
\label{eq:Verdet}
\end{equation}
where $n$ is the refractive index, $C_0$ is a constant that depends on $t$ and the specific form of $Q_{\alpha}({\bf k})$ (see also the companion paper~\cite{Volosniev2022}). First of all, we note that in the limit $\mu\!\!=\!\!0$ (pure minimal coupling to Eq.~(\ref{eq:naive})) $V(\omega)\!\! \sim \!\! 1/\left( \Delta^2-\omega^2\right)^2$, which qualitatively disagrees with our experimental data presented in Fig.~\ref{fig:faraday}B. In the opposite limit ($t=0$), $C_0=0$, and hence $V(\omega\!=\!0)\!=\!0$. Therefore, the observed functional form of $V(\omega)$ can be explained only if both $\mu$ and $t$ are finite. It is also noteworthy that around $\omega=0$, the Becquerel curve computed for the data in Fig.~\ref{fig:fit_mu}B and $\gamma$ chosen to match the high frequency data~\citep{Sabatini2020}, differs from the fit according to Eq.~(\ref{eq:fit_Verdet}) by only a constant offset, see Fig.~\ref{fig:faraday}B.
 
  \begin{figure}
\includegraphics[scale=0.195]{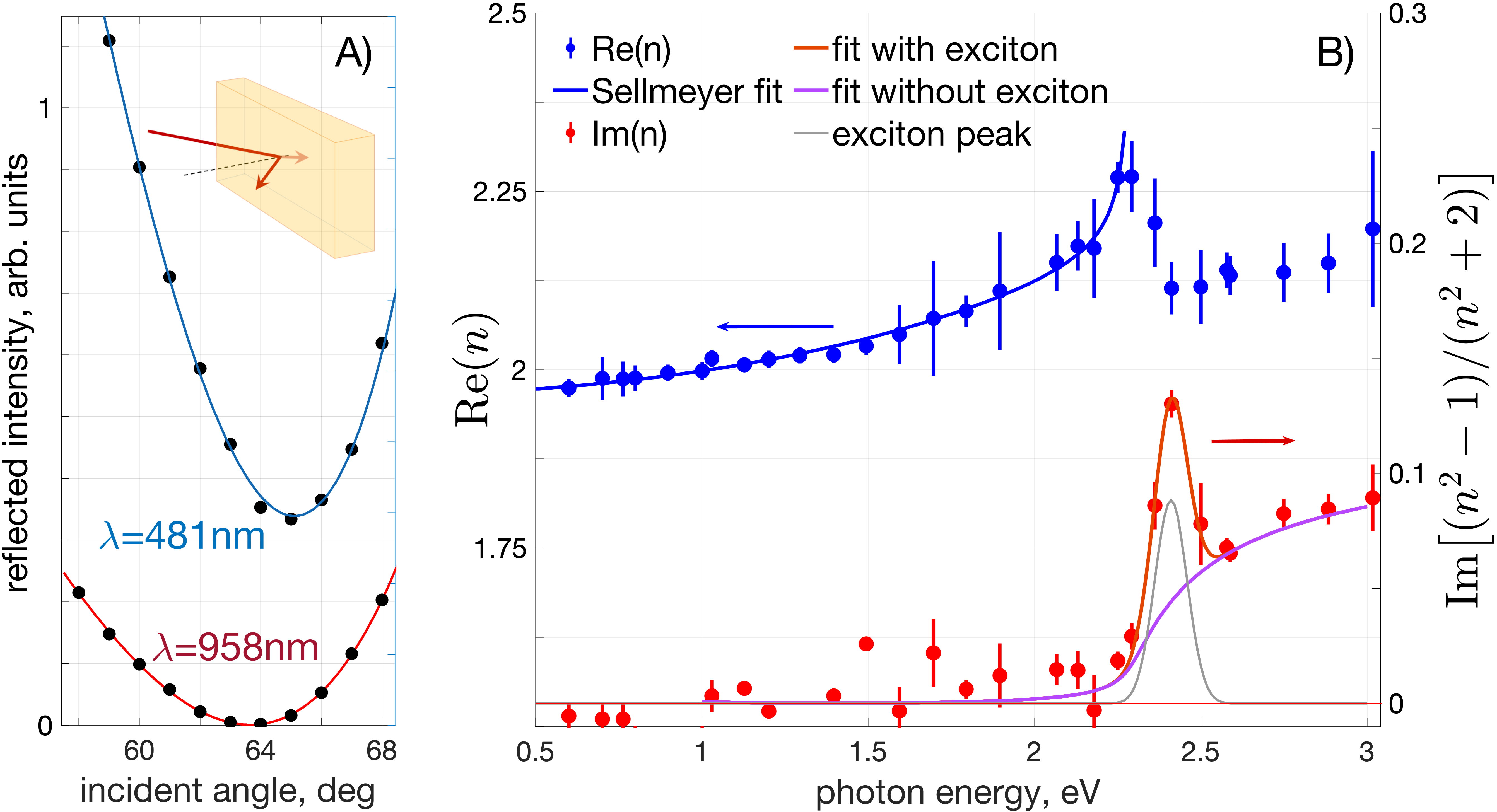}
\caption{A) Reflectivity (p-polarization) as a function of the angle of incidence for two different wavelengths with Fresnel fits to extract the complex refractive index. Note that the reflectivity has a non-zero minimum value for $\hbar \omega > \Delta$ implying non-zero imaginary part of $n(\omega)$. Inset: Brewster angle measurement. B) Blue: Experimental values of Re$\left( n \right)$ (dots) plotted together with a Sellmeyer fit (solid curve); Red: Experimental values of Im($\alpha$) (dots) extracted from the imaginary part of the refractive index plotted together with our fit (solid curve). The purple curve is a fit based on our effective model (without excitons). The grey curve shows the excitonic peak. Note that both $t$ and $\mu$ terms are necessary for a faithful fit, see the Supplementary Material~\cite{SM}.}
\label{fig:fit_mu}
\end{figure}

{\it Determination of $\mu$ and $\tilde \mu_B$.---} The spin-electric term is essential for the observed frequency dependence of the normalized Verdet constant $\tilde{V}(\omega)$. However, the Faraday effect alone does not allow us to estimate the value of $\mu$, since $\mu$ enters the corresponding part of Eq.~(\ref{eq:Verdet}) as a product with the unknown magnetic moment $\tilde \mu_B$. Therefore, we measure the complex refractive index. This quantity has been previously measured on micro-scale crystals CH$_3$NH$_3$PbBr$_3$~\cite{Brittman2016}. In order to avoid possible effects due to the differences in sample geometries, we perform a complementary analysis of the refractive index $n(\omega)$ in a bulk single crystal CH$_3$NH$_3$PbBr$_3$ by measuring the reflection of the beam as a function of the angle of incidence near the Brewster minimum. By fitting the angle dependence using the Fresnel expressions~\cite{Zommerfeld1954}, one can extract $n$. This is similar to the previous experiment~\cite{He2019}, however we also go into shorter wavelengths ($\hbar \omega > \Delta$). In this regime the refractive index becomes complex, $n(\omega) = n+i\,k$. The imaginary part $k$ manifests itself as a finite minimal reflectivity (see Fig.~\ref{fig:fit_mu}, and the Supplementary Material~\cite{SM}). 

Complex-valued $n$ can be connected to the polarizability $\alpha(\omega)$ via the Lorentz-Lorenz relation (see, e.g.,~\cite{Boyd2008,FeynmanVol2}). For the purpose of determining the low-energy parameters, we are most interested in the imaginary part of $\alpha(\omega)$ since, unlike the real part, it is ``local'' in frequency and does not receive contributions from the high-energy degrees of freedom. This quantity can be readily calculated based on Eq.~(\ref{eq:Hamiltonian}) (see the companion paper~\cite{Volosniev2022}). To extract the value of $\mu$ by fitting the experimental data in Fig.~\ref{fig:fit_mu}, we first need to estimate the other parameters that enter Eq.~(\ref{eq:Hamiltonian}): $\Delta \!\! \approx \!\! 2.3$eV is known from optical measurements (see, e.g., Refs.~\cite{Pazhuk1981, Saidaminov2015, Jacobsson2016,Ng2018}, also see Fig.~\ref{fig:fit_mu}); $t\!\simeq \! 0.6$eV and the bandwidth $t_3\! \simeq\! 0.9$eV of $\epsilon({\bf k})$ can be estimated by comparing the low-energy electron dispersion (see Eq.~(\ref{eq:energy_low_k})) to the first-principles numerical calculations of Ref.~\cite{Becker2018} where $a=0.586$nm~\footnote{The parameters in this work were calculated for CsPbBr$_3$ which has a very similar lattice to that of CH$_3$NH$_3$PbBr$_3$; $t$ was determined from the Kane energy, $E_p$: $t^2=E_p/(24 m_0 a^2)$, where $m_0$ is the free-space electron mass. The parameter $t_3$ was determined by the parameters $\gamma_{e,h}$: $t_3=(\gamma_e+\gamma_h)/(a^2m_0)$.}. With these values, fitting of the polarizability gives $\mu\!\simeq \!0.29 q a$ (Fig.~\ref{fig:fit_mu}) in line with the natural estimate above; then $\tilde \mu_B \! \simeq \!-0.4 \mu_{B}$ (Bohr magnetons). Importantly, the particular form of $\epsilon ({\bf k})$ and $Q_{\alpha}({\bf k})$ does not affect the value of $\mu$ noticeably (see the companion paper~\cite{Volosniev2022}); the value of $\tilde \mu_B$ is however sensitive to the band structure, and we can only estimate it. 

{\it Discussion.---} Now that we have established that the spin-electric coupling is essential to describe response of the material to electromagnetic fields, we relate the Hamiltonian in Eq~(\ref{eq:Hamiltonian})
  to basic symmetries of the system. For the considered CH$_3$NH$_3$PbBr$_3$, these are the cubic point group $O_h$~\footnote{Note that here we are using the Sch{\"o}nflies notation; in the Hermann–Mauguin notation the point group would be $m3m$.} which includes inversion $\hat{I}$, and time-reversal symmetry~$\hat{T}$:
\begin{equation}
\hat{T}= i\tau_3\otimes \sigma_2 \hat{K}, \qquad \hat{I}=-\tau_3\otimes\sigma_0,
\label{eq:symmetries}
\end{equation}
where $\hat{K}$ is the complex conjugation operator. This set of symmetries is quite restrictive and it turns out that Eq.~(\ref{eq:Hamiltonian}), with $\epsilon({\bf k})$ and $Q_{\alpha}({\bf k})$ being even and odd functions of ${\bf k}$ respectively, contains all basis coupling terms for the states of Eq.~(\ref{eq:states}) up to linear order in $\vec{E}$ and $\vec{B}$~\footnote{
Note that any product of the terms in Eq.~(\ref{eq:Hamiltonian}), e.g., $\tau_3\!\otimes\!\left[\vec{\sigma}\!\cdot\!\left(\vec{k}\!\times\!\vec{E}\right)\right]$, obey the required symmetries. Therefore, these terms could, in principle, be also added to the effective Hamiltonian. Being the products of the basis coupling terms, they do not lead to any new dependence of observables on the parameters of the system and are not important for our discussion. We leave their investigation to future studies, see also the companion paper~\cite{Volosniev2022}.}.

Identification of the explicit form of $H$ is the central theoretical result of this Letter. Below, we shall briefly discuss its physical implications. In the most simple situation with no external fields the  electron dispersion near $k=0$ can be found by simply taking square of $H_k$ from Eq.~(\ref{eq:naive})
\begin{equation}
\mathcal{E}_k^2=\frac{1}{4}\left(\Delta+t_3(\mathbf{k}a)^2\right)^2+(2t\,ka)^2.
\end{equation} 
The energy levels are doubly degenerate, and can be parametrized by the spin degree of freedom -- the direction of (quasi-)spin is determined by the direction of the momentum $\mathbf{k}$. The degeneracy is broken when external electric field $\vec{E}$ is applied. This can be seen by squaring the Hamiltonian with $B=0$:
\begin{equation}
H^2\!=\!\left( \mathcal{E}^2_k + \mu^2|\vec{E}|^2 \right)\!+4a t\mu \tau_3\!\otimes\!\left[\vec{\sigma}\!\cdot\!\left(\vec{k}\!\times\!\vec{E}\right)\right].
\label{eq:energy_low_k}
\end{equation} 

\begin{figure}
\includegraphics[scale=0.25]{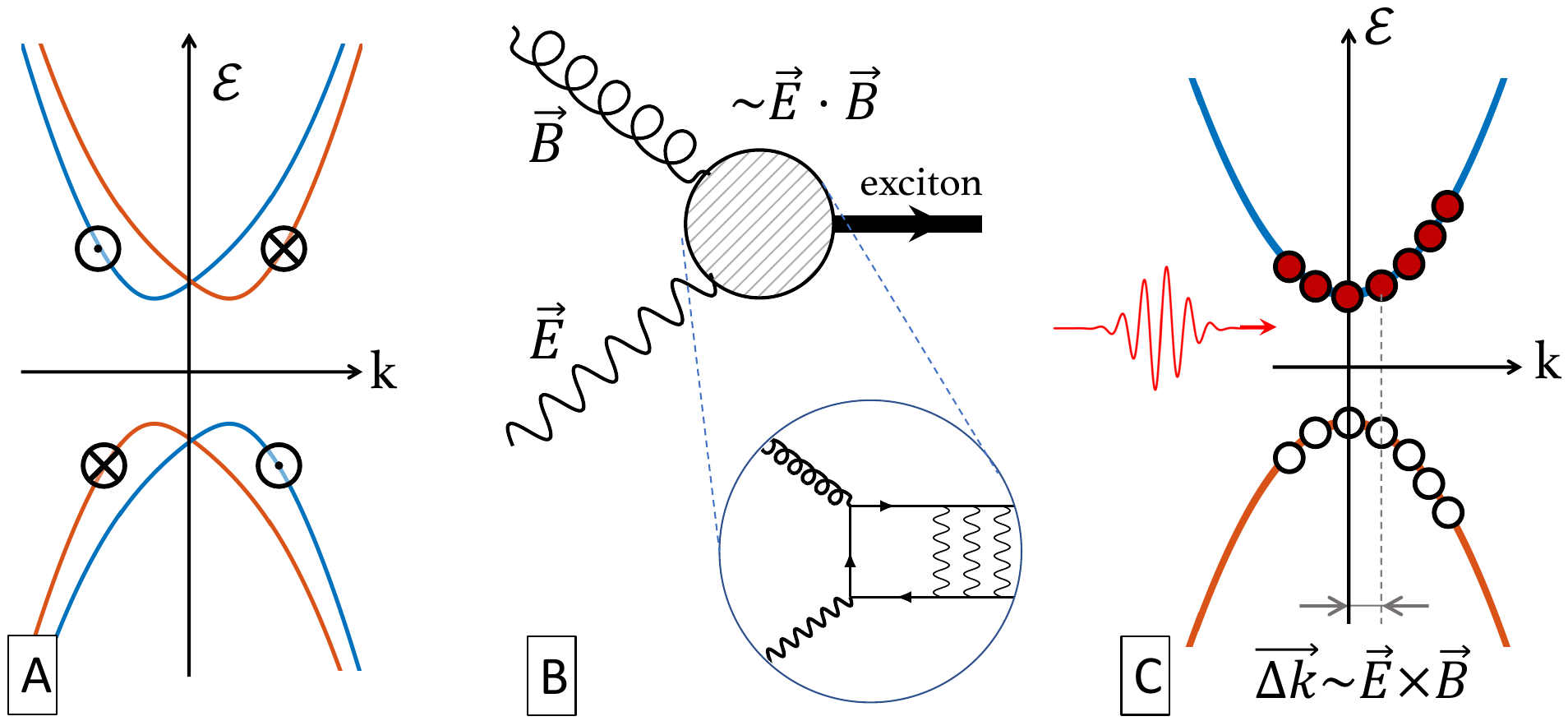}
\caption{A) Rashba splitting that follows from the interplay of the hopping and spin-electric terms; B) Axion-type ($\vec{E}\cdot\vec{B}$) term as a result of the spin-electric and regular Zeeman terms; C) Photo-induced momentum shift under circularly polarized irradiation as a result of the spin-electric and orbital-selective Zeeman terms.}
\label{fig:implications}
\end{figure}

\noindent The degeneracy is indeed lifted and the splitting has a clear Rashba-type form (see Fig.~\ref{fig:implications}A). Notice that both the conduction and valence bands are split, in line with both experiment~\cite{Niesner2016} and {\it ab initio} theory~\cite{Kim2014}. This equation also indicates that the gap in the spectrum is modified in the presence of electric fields: $\delta\Delta_{gap}\propto E^2$, which is consistent with the observed shift of the absorption edge in perovskite samples pumped by THz pulses~\cite{Kim2017}.

In the more complex case of finite $\vec{E}$ and $\vec{B}$, mixing between $H_E$ and $H_B$ can give rise to magneto-electric phenomena taking the following forms:
\begin{equation}
\delta H_{ax}\sim \tau_1\!\otimes\!\sigma_0 \left( \!\vec{E}\!\cdot\!\vec{B} \right),\,\,\,
\delta H_{pd}\sim \tau_2\!\otimes\!\vec{\sigma}\!\cdot\!\left[ \vec{E}\! \times \! \vec{B} \right].
\end{equation}
Both terms are of the order of $\sim\!\!(\tilde \mu_B B)(\mu E)/\Delta$ for $\omega\ll \Delta$. As the structure suggests, $\delta H_{ax}$ describes what is known in high-energy physics as axion electrodynamics~\cite{Sekine2021}. In the case of perovskites the role of the axionic field $\phi_{ax}$ is played by a certain exciton (Fig.~\ref{fig:implications}B)~\footnote{\unexpanded{$\phi_{ax} \sim \langle \tau_1^{\alpha \beta} \delta^{a b} c^{\dagger}_{\alpha a} c_{\beta b} \rangle$}, where $c^{\dagger}_{\alpha a}$ ($c^{\dagger}_{\alpha a}$) is a fermionic creation (annihilation) operator.}. $\delta H_{pd}$ corresponds to photon drag. Indeed, comparing it with $H_k$ one can see that it gives rise to a shift in ${\bf k}$-space in the direction of the Poynting vector of the incident beam. This should manifest itself as a photo-current in a photoexcited perovskite sample (Fig.~\ref{fig:implications}C). Unlike the usual photon drag mechanism that diverges at $\omega \rightarrow 0$~\cite{Wegener}, $\delta H_{pd}$  should resonantly diverge at $\omega \rightarrow \Delta$.  Finally, $\tau_2\otimes \vec{\sigma}$ changes sign under inversion and time-reversal and can be interpreted as a toroidal moment~\cite{DUBOVIK1990}. Therefore, LHP in perpendicular magnetic and electric fields can be used to study ferrotoroidicity and related effects~\cite{Spaldin2008}\footnote{Note that analysis of corresponding experimental data will benefit from the recent theoretical studies of the magnetoelectric effect, see, e.g., Refs.~\cite{Spaldin2013,Gao2018,Thole2020}}.

In conclusion, we have analyzed the Faraday effect in CH$_3$NH$_3$PbBr$_3$ in a broad wavelength range. We find that to describe the observed frequency dependence of the Verdet constant one needs to consider the local atomic response to applied electric fields. We also find that the coupling has a specific ``spin-electric'' form. This allows us to quantitatively describe our measurements.  
The existence of multiple channels for electromagnetic coupling, combined with exceptional polarizability of Pb$^{2+}$ ions is shown to give rise to a rich variety of novel physics in LHP including anomalous photon drag and axion electrodynamics.

Our theoretical model is constructed for valence and conduction bands made of s-type and p-type doublets in centrosymmetric cubic materials with strong spin-orbit coupling. This suggests further materials where the ``spin-electric'' coupling can be observed, such as cubic tin-based halide perovskites (e.g., CsSnBr$_3$)\cite{Borriello2008,Huang2013} with tin in a highly polarizable Sn$^{2+}$ state~\cite{Schwerdtfeger2019}.

\begin{acknowledgments}
We thank Maksym Serbyn, Areg Ghazaryan and Nuh Gedik for useful discussions; M.L. acknowledges support  by the European Research Council (ERC) Starting Grant No. 801770 (ANGULON).
\end{acknowledgments}


\clearpage


\section{Sample Preparation}
\noindent{\it Chemicals.---}CH$_3$NH$_3$Br ($>$99.99\%) was purchased from GreatCell Solar Ltd. (formerly Dyesol) and used as received. PbBr$_2$  ($\geq$98\%),  CsBr  (99.9\%  trace  metals  basis),  DMF  (anhydrous,  99.8\%),  and  DMSO  (anhydrous, $\geq$99.9\%)  were  purchased  from  Sigma Aldrich  and  used  as  received. 

\noindent{\it Synthesis of CH$_3$NH$_3$PbBr$_3$ perovskite single crystals.---}A 1.5 M solution of CH$_3$NH$_3$Br/PbBr$_2$ in DMF was prepared, filtered through a 0.45-$\mu$m-pore-size PTFE filter; and the vial containing $0.5-1$ ml of the solution was placed on a hot plate at $30^{\circ}$C. Then the solution was gradually heated to $\sim\!60^{\circ}$C and kept at this temperature until the formation of CH$_3$NH$_3$PbBr$_3$ crystals. The crystals can be grown into larger sizes by elevating the temperature further. Finally, the crystals were collected and cleaned using a Kimwipe paper.

\section{Faraday rotation measurement}

\begin{figure}
\includegraphics[scale=0.27]{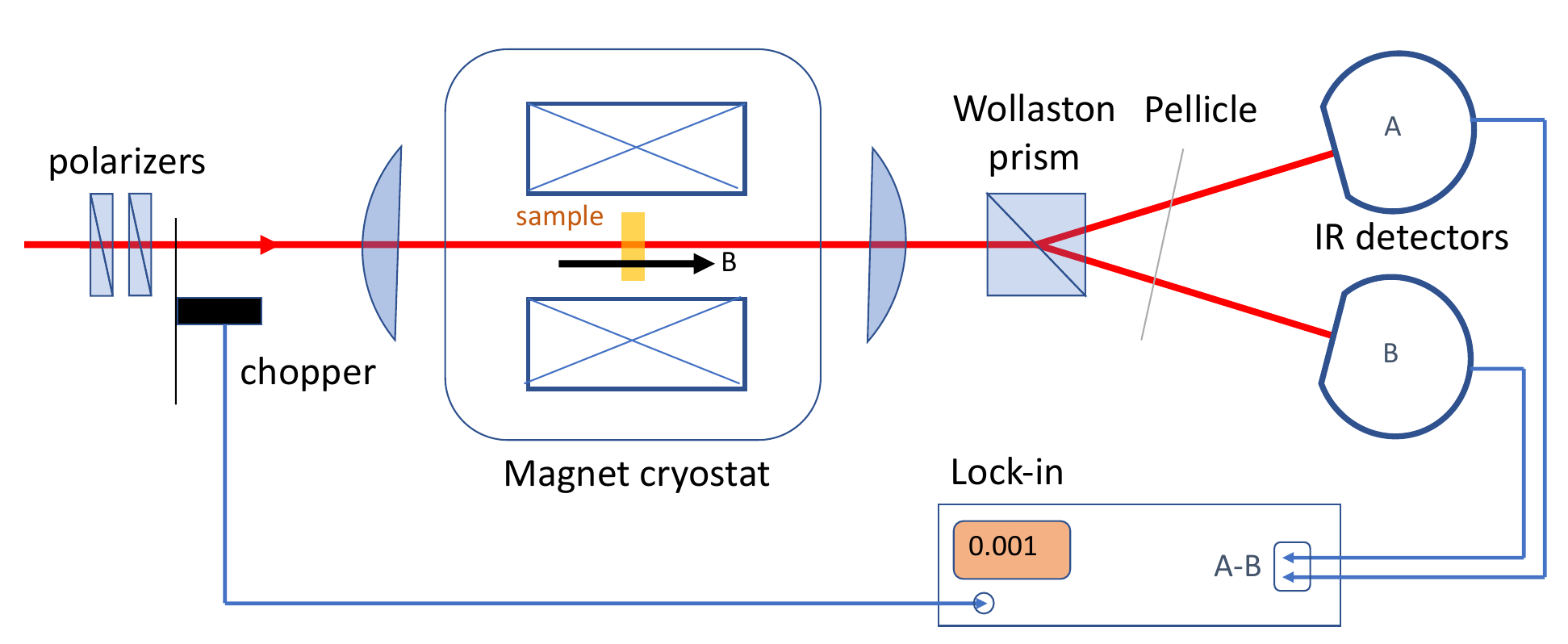}
\caption{Faraday rotation measurement setup. The sample is kept inside an optical magnet cryostat; the polarization rotation is detected with a pair of balanced infra-red detectors.}
\label{fig:setup}
\end{figure}

We are measuring the Faraday rotation away from the  band edge, therefore the net value of the polarization rotation is not very large. For that reason, we are not measuring the Faraday angle directly with a polariser as in~\cite{Sabatini2020}. Instead, we infer it from a balanced detection scheme as shown in Fig.~\ref{fig:setup}. Here, the beam is first passing through a pair of polarizers whose purpose is two-fold. On the one hand, it cleans the polarization. On the other hand, it allows for an intensity control of the probing beam. Next, the beam is focused with an f=200mm lens onto the perovskite sample which is kept in vacuum inside an optical magnetic cryostat ({\it Oxford Instruments} SpectromagPT). The beam is then recollimated on the other side of the cryostat. The outer windows of the cryostat are made of ZnSe while the inner ones are CVD-grown diamond, both are transparent in the relevant IR range.

The outgoing collimated beam passes through a Wollaston prism whose two outputs are send to two pyroelectric IR detectors ({\it Gentech} THZ5I-BL-BNC). To compensate for the inevitable difference in sensitivities of the detectors, we introduce a pellicle beam splitter after the Wollaston prism, which can be rotated to achieve different transmissivities for the two cross-polarized Wollaston prism outputs.

CH$_3$NH$_3$PbBr$_3$ is considered to be nominally optically isotropic in the high-temperature ($T>T_c=240$K) cubic phase. Nevertheless, the single-crystal samples are in general birefringent even above $T_c$, presumably due to built-in stress. For this reason, for a generic incoming linear polarization the outgoing polarization will be elliptical, complicating the interpretation of the data. To avoid this, we adjust the polarization of the probe light to be along the optical axis of the sample. These happen to be parallel to the edges of CH$_3$NH$_3$PbBr$_3$ crystals. For precise alignment we minimize the ellipticity of the outgoing beam as a function of incoming polarization orientation. Next, in order to use the balanced detection scheme for measuring the changes in polarization, the Wollaston prism has to be oriented at 45$^{\circ}$ relative to the unperturbed polarization direction. Typically this is achieved by placing a $\lambda/2$ waveplate in front of the Wollaston prism. However, in our case this is not practical as we need to vary the wavelength is a broad range. Therefore we actually rotate the Wollaston prism by $45^{\circ}$ from the position where the cross-polarized output is minimal. 

To balance the detector sensitivities, we place an additional infrared polarizer in front of the Wollaston prism to clean the polarization (wire grid on ZnSe. The advantage of this option over, e.g., Glan-Taylor prism, is that the former is less sensitive to the incoming direction and does not change the outgoing beam path). This polarizer is set to be parallel to the nominal polarization of the beam outgoing from the sample and hence 45$^{\circ}$ relative to the Wollaston prism. This way we know that both output beams from the Wollaston have the same intensity and any difference in the readings of the detectors comes from their different sensitivities which can be compensated for by rotating the pellicle beam splitter.

Once the detectors are balanced, their outputs $s_1$ and $s_2$ are sent to the differential inputs of a lock-in ({\it SRS} SR830) synchronized with an optical shopper modulating the input intensity. A simple Jones analysis shows that the Faraday rotation angle can be found as

\begin{equation}
\sin\left(2\Theta_F\right)=\frac{s_1-s_2}{s_1+s_2}.
\label{eq:faraday}
\end{equation}    

\begin{figure}
\includegraphics[scale=0.4]{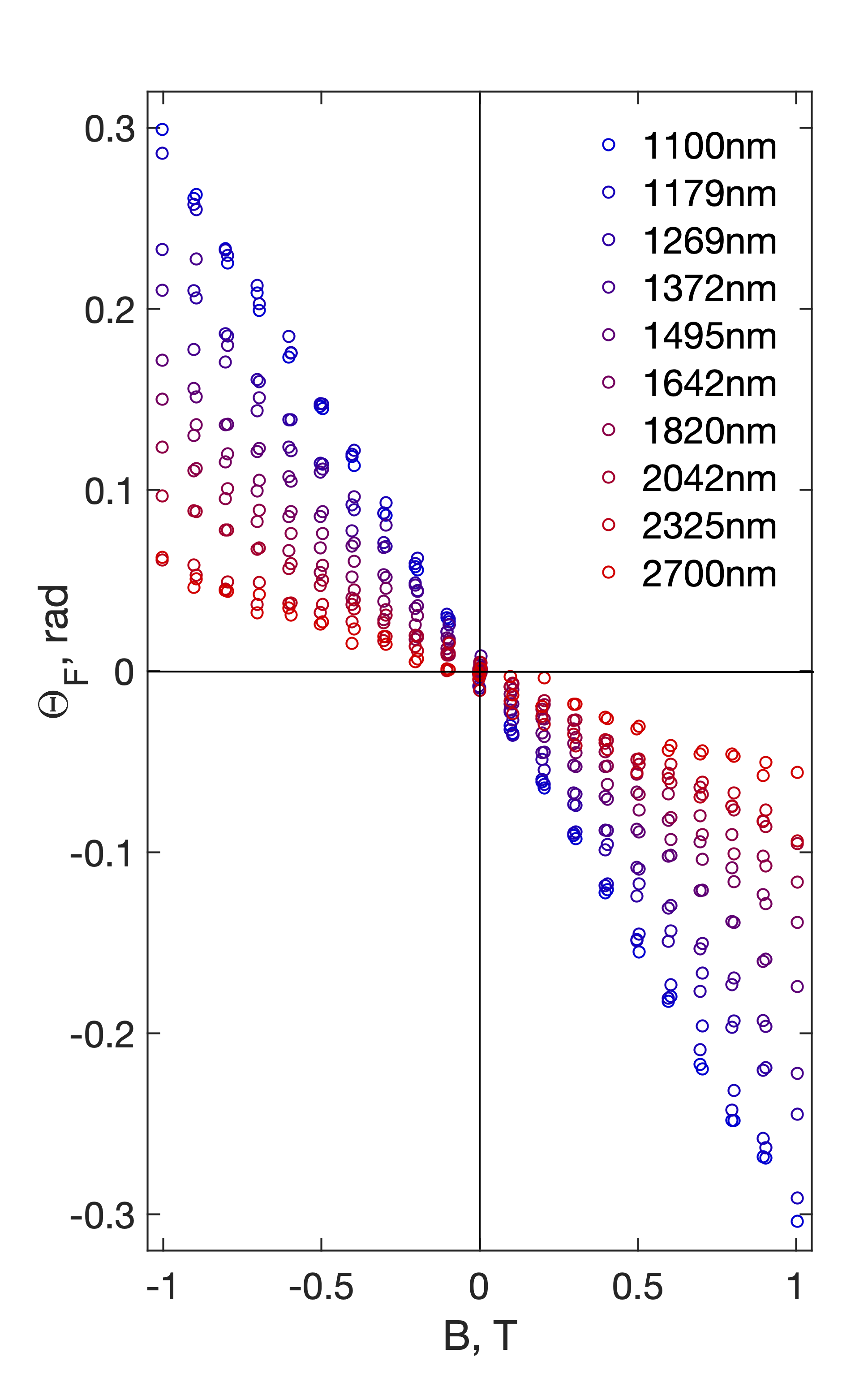}
\caption{Faraday rotation $\Theta_F$ as a function of the magnetic field $B$.}
\label{fig:faraday_SM}
\end{figure}

\noindent As we know the thickness of the sample ($d\!\approx\!1.66$mm), the Verdet constant can be now deduced from the slope of $\Theta_F(B)$, see Fig.~\ref{fig:faraday_SM}.

\begin{figure}
\includegraphics[scale=0.5]{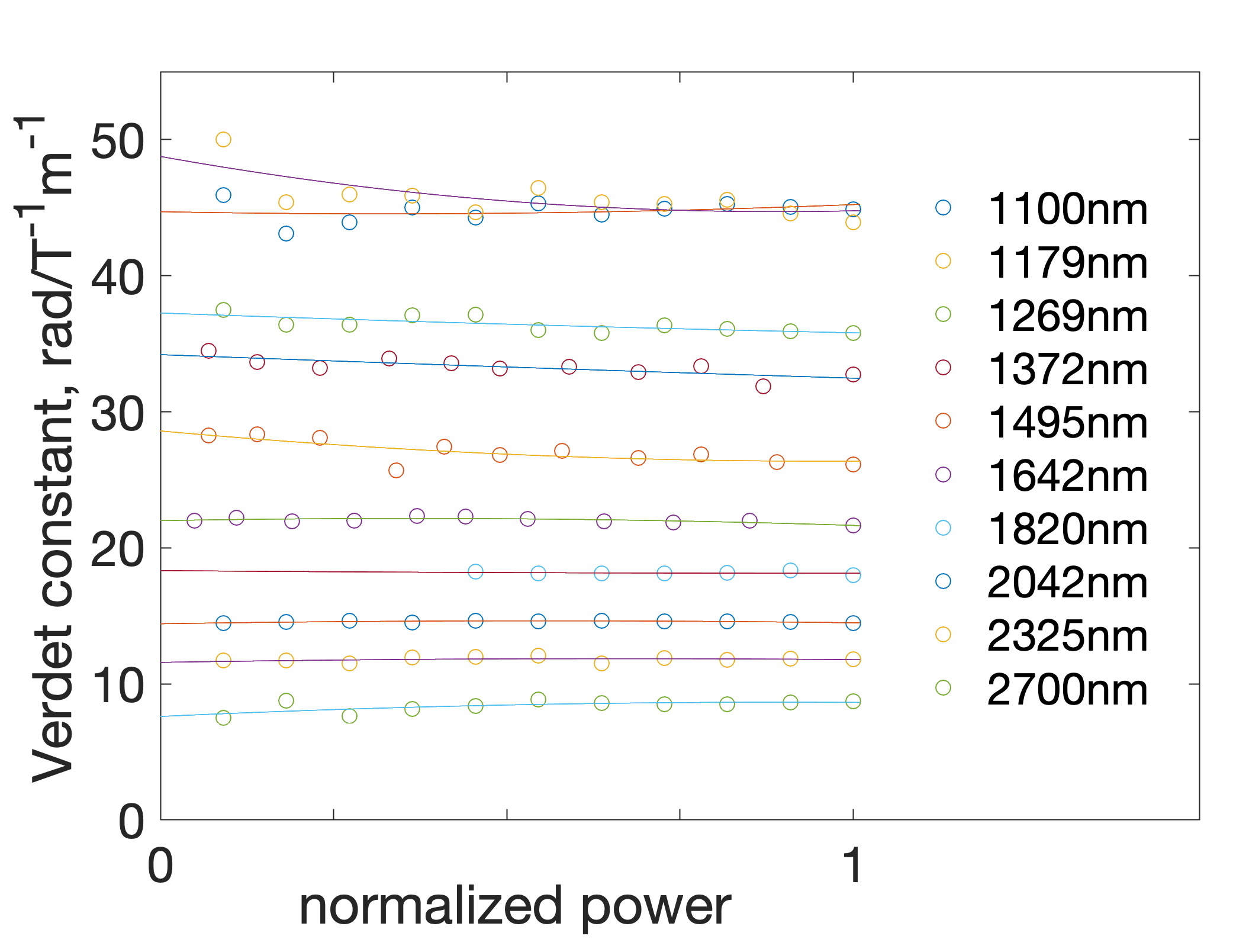}
\caption{Verdet constant as a function of the intensity of the incident probe beam at different wavelength values. Solid lines are the second-order polynomial fits used to extrapolate the data to zero intensity.}
\label{fig:power}
\end{figure}

\noindent {\it Light source and intensity dependence.---}The variable wavelength (1100nm-2700nm) probe beam is generated by an Optical Parametric Amplifier ({\it Light Conversion} OrpheusHP), pumped by a amplified pulsed laser ({\it Light Conversion} Pharos; $\lambda\!=\!1028$nm, 290fs pulse duration, 6W@3kHz repetition rate). The typical power of the probe beam  at the entrance window of the cryostat was a few mW. In order to avoid possible non-linear effects, for each wavelength we have measured the Faraday rotation at different power values and extrapolated the observed numbers to the zero intensity, as plotted in Fig.~{\ref{fig:power}}. As can be seen the value of the Verdet constant is almost independent of the power of the incident beam, with an exception of shorter wavelengths $\lambda\simeq1\mu$m. We believe this is due to enhanced nonlinearity near the two-photon absorption threshold $\lambda=530nm$ \cite{SheikBahae1990}. The error bars in Fig.1B of the main text are the 95\% confidence range of the interpolated Verdet values.  

\begin{figure}
\includegraphics[scale=0.4]{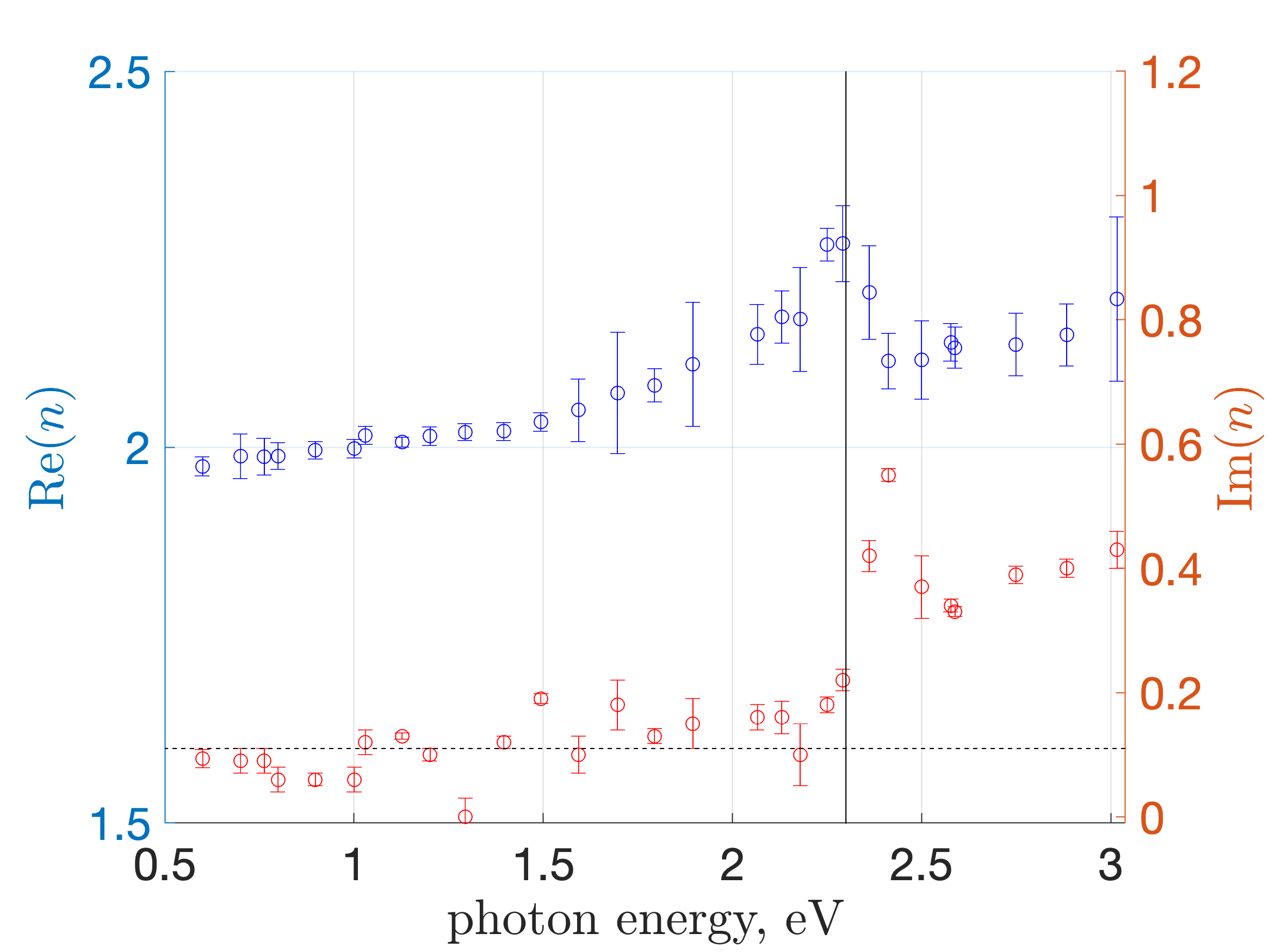}
\caption{Real (blue) and imaginary (red) parts of refractive index as inferred from Brewster angle measurements.}
\label{fig:refractive}
\end{figure}

\section{Refractive index measurement}

\begin{figure}
\includegraphics[scale=0.4]{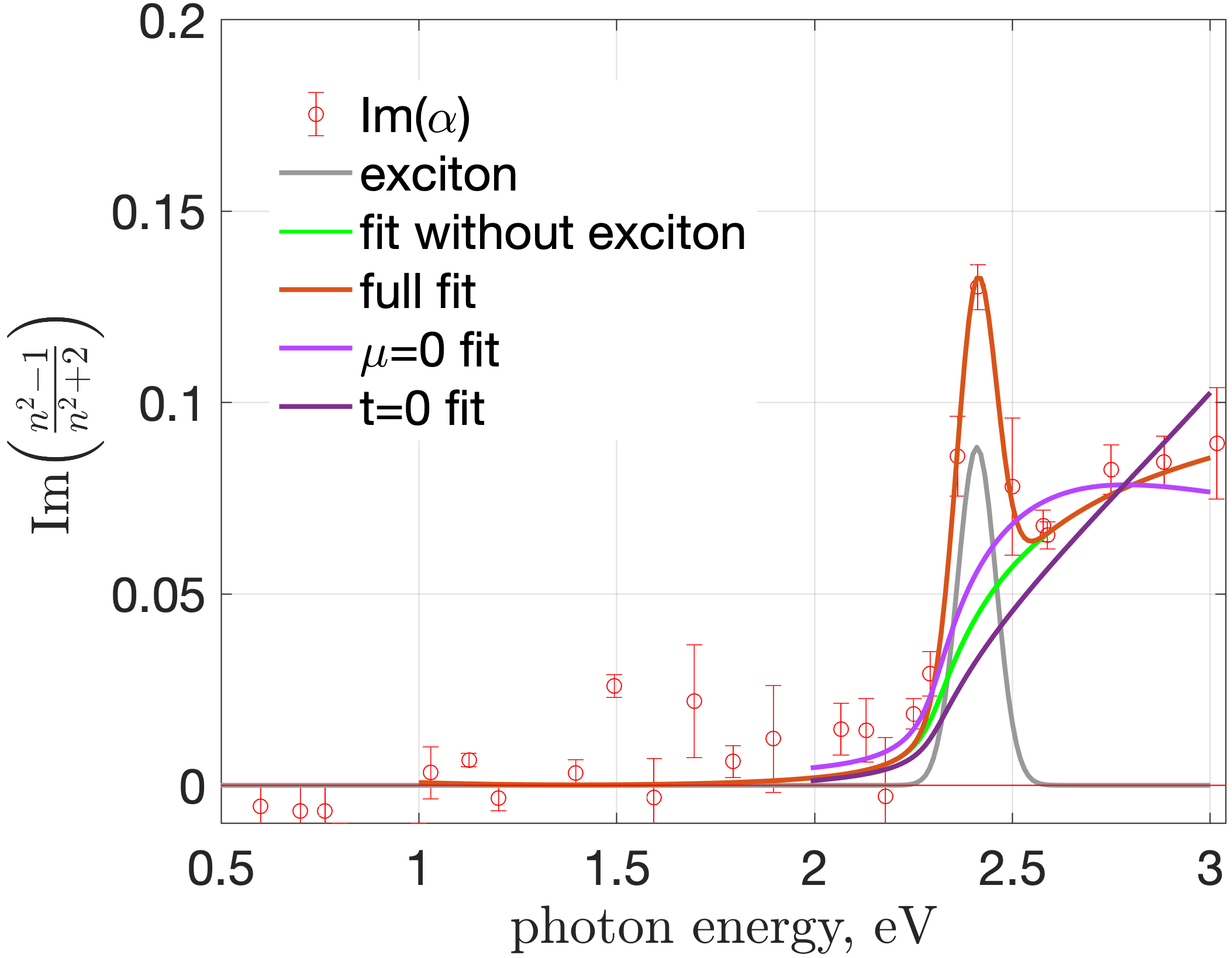}
\caption{Red dots: Imaginary part of the polarizability extracted from the imaginary part of the refractive index, see Fig.~\ref{fig:refractive}; Green curve: fit with $t=0.6$eV ($t$ is fixed here to the expected value~\cite{Becker2018}) and $\mu=0.29qa$; Magenta curve: fit with $t=1.08$eV and $\mu=0$; Purple curve: fit with $t=0$ and $\mu=0.64qa$; Gray: exciton peak Gaussian fit;  Red curve: full fit (sum of the green and gray curves). Note that both $t$ and $\mu$ terms are necessary for a faithful fit.}
\label{fig:fit_mu_SM}
\end{figure}

Refractive index was measured on single crystal CH$_3$NH$_3$PbBr$_3$ samples by measuring the reflectivity of a p-polarized beam off the sample surface. Prior to the measurement the sample surface was polished using a 1$\mu$m-grit Al$_2$O$_3$ sandpaper. The sample was then placed on a rotation stage. For each wavelength value the zero-position was found by finding the backward-reflection configuration. The reflection coefficient $R$ was then measured as a function of the angle of incidence $\theta$. The resulting curves were fit by the Fresnel formula
\begin{equation}
R(\theta)=\left|  \frac{n_a \cos \theta - n_p \sqrt{1-\left( \frac{n_a}{n_p} \sin\theta  \right)}}{n_a \cos \theta + n_p \sqrt{1-\left( \frac{n_a}{n_p} \sin\theta  \right)} } \right| ^2.
\label{eq:fresnel}
\end{equation}    
\noindent Here,  $n_a\approx 1$ and $n_p$ are the phase refractive indices of air and CH$_3$NH$_3$PbBr$_3$ respectively. In order to achieve good fitting in the whole wavelength range, including $\hbar \omega > \Delta$ (the position of the energy gap is marked by the (black) vertical line in Fig.~\ref{fig:refractive}), one needs to assume that the refractive index is a complex number, i.e., $n_p=n+i k$. The resulting measured complex refractive index is shown in Fig.~\ref{fig:refractive}. As can be seen, the value of Im$(n)$ for wavelength values longer than absorption edge is seemingly finite. This reflects the fact that there is a non-zero minimal reflectivity. This can either point towards an actual parasitic absorption inside    CH$_3$NH$_3$PbBr$_3$ due to impurities, but given that the samples ($>\!\!1$mm thick) actually look transparent in the visible range ($\lambda\!>\!530$nm) we attribute this residual Im$(n)$ to the occasional surface imperfections that scatter incoming light. Therefore, we take this level as an offset due extrinsic effects (the (black) horizontal line in Fig.~\ref{fig:refractive}). We subtract it from the data for all further usage. The fit to the data is presented in the main matter as well as in Fig.~\ref{fig:fit_mu_SM}.

\bibliography{refs}
\end{document}